\documentclass[12pt]{article} 

\usepackage[dvips]{graphicx}
\usepackage{multicol}
\usepackage{epsfig}
\usepackage{float}
\usepackage{bm}        
\usepackage{amssymb}   
\usepackage{amsmath}
\usepackage{framed}
\usepackage{xcolor}
\definecolor{shadecolor}{named}{gray}
\usepackage{paralist}
\usepackage{enumitem}
\usepackage{lscape}
\usepackage{caption}
\usepackage{tabularx}
\newcolumntype{Y}{>{\centering\arraybackslash}X}
\usepackage{wrapfig}
\usepackage[font=scriptsize]{caption}
\usepackage[square,numbers,compress]{natbib}
\usepackage{doi}
\usepackage{soul}
\usepackage{ulem}

\newcommand{\LL}[2]{\Lambda_{#1, #2}}

\usepackage{authblk}
\usepackage{blindtext}
\usepackage{hyperref}
\begin{document}
\title{\large  Optical Atomic Clock aboard an Earth-orbiting Space Station (OACESS): \\
Enhancing searches for physics beyond the standard model in space \\\vspace{2mm}
}
\author[1]{Vladimir Schkolnik}
\author[2,3]{Dmitry Budker}
\author[1]{Oliver Fartmann}
\author[4]{Victor Flambaum}
\author[5]{Leo Hollberg}
\author[6]{Tigran Kalaydzhyan}
\author[7]{Shimon Kolkowitz}
\author[1]{Markus Krutzik}
\author[8]{Andrew Ludlow}
\author[8]{Nathan Newbury}
\author[1]{Christoph Pyrlik}
\author[8]{Laura Sinclair}
\author[9,10]{Yevgeny Stadnik}
\author[1]{Ingmari Tietje}
\author[11]{Jun Ye}
\author[12]{Jason Williams}
\affil[*]{\small Primary Author: \textbf{vladimir.schkolnik@physik.hu-berlin.de}, \textbf{+49 (0)30 2093-7625} }
\affil[1]{\small Humboldt-Universit\"at zu Berlin, Newtonstr. 15, 12489 Berlin, Germany}
\affil[2]{Helmholtz-Institut Mainz, Johannes Gutenberg-Universitat Mainz, 55128 Mainz, Germany}
\affil[3]{Physics Department, University of California, Berkeley 94720-7300}
\affil[4]{School of Physics, University of New South Wales, Sydney 2052, Australia}
\affil[5]{Department of Physics, HEPL, Stanford University, 452 Lomita Mall, Stanford, California 94305}
\affil[6]{ Atomic Developers, 2501 Buffalo Gap Rd \#5933, Abilene, Texas 79605}
\affil[7]{Department of Physics, University of Wisconsin-Madison, Madison, WI 53706}
\affil[8]{National Institute of Standards and Technology, Boulder, Colorado 80305}
\affil[9]{Kavli Institute for the Physics and Mathematics of the Universe (WPI), The University of Tokyo Institutes for Advanced Study, The University of Tokyo, Kashiwa, Chiba 277-8583, Japan}
\affil[10]{School of Physics, The University of Sydney, NSW 2006, Australia}
\affil[11]{JILA, National Institute of Standards and Technology and University of Colorado Department of Physics, University of Colorado Boulder, Colorado 80309-0440}
\affil[12]{Jet Propulsion Laboratory, California Institute of Technology, Pasadena, CA 91109}

\maketitle

\pagebreak

\begin{abstract}

We present a concept for a high-precision optical atomic clock (OAC) operating on an Earth-orbiting space station. This pathfinder science mission will compare the space-based OAC with one or more ultra-stable terrestrial OACs to search for space-time-dependent signatures of dark scalar fields that manifest as anomalies in the relative frequencies of station-based and ground-based clocks. This opens the possibility of probing models of new physics that are inaccessible to purely ground-based OAC experiments where a dark scalar field may potentially be strongly screened near Earth's surface. This unique enhancement of sensitivity to potential dark matter candidates harnesses the potential of space-based OACs. 

\end{abstract}

\pagenumbering{arabic}

\section{Introduction}

Clocks play a critical role in communications technology, synchronization of large data networks and radio telescopes, and in generating reference signals for global satellite navigation systems. 
Recent advances in optical atomic clocks (OAC) provide new measurement capabilities with which the Standard Model (SM) and General Relativity (GR) can be tested to unprecedented precision \cite{Safronova2018,antypas2022new,Bothwell2022,Zheng2022}. These clocks utilize ensembles of laser-cooled atoms as frequency references. The clock signal is derived from an ultra-stable laser which is frequency-locked to a narrow atomic  transition at optical frequencies of 100's of THz \cite{Matei2017,Zhang2017}. Terrestrial OAC have now achieved stability and accuracy at the 10$^{-18}$ level in fractional frequency units \cite{Bothwell2022,RevModPhys.87.637,Campbell2017,McGrew2018,Schioppo2017,Oelker2019,Hinkley2013,Marti2018,Bothwell2019}, orders of magnitude better than the record caesium microwave frequency standards \cite{BACON2021}. It is envisioned that future precision OAC operating in Earth's orbit will serve as invaluable infrastructure for a global quantum network of clocks with enabling capabilities for a range of scientific and technical applications \cite{Komar2104}. 
Future generations of global navigation satellite systems (GNSSs) consisting of advanced OACs could be used also for deep space navigation in our solar system. In addition, such a system would generate scientific data for dark matter searches at the same time
\cite{Schuldt:2021}.

It has recently been realized that space-based clocks should offer a significant advantage over ground-based clocks to test certain models of dark-sector particles, fields, or entities (such as dark matter and dark energy) which are absent in the SM; specifically, for dark scalar fields where the scalar field is strongly screened near Earth's surface \cite{Hees:2018screen,Stadnik:2020screen,Stadnik:2021screen} (see also \cite{Damour:1993screen,Pospelov:2008screen,Khoury:2010screen} for earlier work on screening in scalar-field models). 
The Optical Atomic Clock aboard an Earth-orbiting Space Station (OACESS) 
concept therefore addresses two priorities for NASA \cite{Decadal2021}: 1) By establishing high-precision OAC and ground-to-space time-transfer technologies in Earth's orbit, it provides capabilities for science that could profit strongly if done in space,
with anticipated value to humans on Earth, and 2) The possibility of detecting dark fields which are screened near Earth's surface enables scientific experiments that, with current technologies, can only be done in space.

We propose a strontium OAC operating aboard an Earth-orbiting space station as a pathfinder science mission to enable unprecedented capabilities in searches for signatures of physics beyond the SM. Comparison of the OACESS clock with terrestrial OAC via leading-edge optical time-frequency transfer techniques \cite{BACON2021,Bodine2020, Hollberg2022, Berceau2016, Robert2016, Taylor2017, Taylor2020} will enable tests of temporal and spatial variations of the fundamental constants of nature that can arise in various dark-sector models, including ultra-low-mass dark matter and dark energy, as well as in quantum gravity. In this way, OACESS will serve as a pathfinder for dedicated missions (e.g., FOCOS \cite {FOCOS2021}) to establish high-precision OAC as space-time references in space. 

As a space-station-based multi-user research campaign, we also consider the inclusion of high-precision sensors for complementary searches for dark-matter signatures, including magnetometers, accelerometers, optical cavities, and interferometers. Additional sensors or measurement combinations allow for complementary searches for dark-matter signatures. Vapor-cell-based magnetometers are routinely used in both space mission and ground based applications \cite{Acuna2002,Korth2016}. Mature laser technology, microfabricated cells, and optical assembly techniques also allow for a compact low Size, Weight, and Power (SWaP) piggyback device.

\section{Probing the dark sector with space-based clocks}
\label{Sec:Theory}

Determining the properties of dark matter and dark energy is one of the grand challenges of our time. 
Ultra-low-mass (sub-eV) bosons are excellent candidates to explain the dark matter and dark energy, and can also serve as more general dark-sector components that do not necessarily play a significant role in cosmology. 
Atomic clocks have been proposed as sensitive probes of dark scalar (spin-0) fields that induce temporal or spatial variations of the fundamental constants of nature, including in models of dark matter \cite{Arvanitaki:2015DM,Stadnik:2015DM1,Stadnik:2015DM2,Hees:2018screen}, solitons \cite{Derevianko:2014solitons,Stadnik:2020screen}, and bursts of relativistic scalar particles \cite{Dailey:2021multi,Stadnik:2021screen}.

Let us consider the following interaction Lagrangian describing the coupling of a real scalar field $\phi$ to ordinary matter at low energies (see, e.g., \cite{Stadnik:2015DM1,Stadnik:2015DM2}): 
\begin{align}
\label{scalar-field_Lagrangian}
 \mathcal{L}^{(n)}_\textrm{int} = \left(\frac{\phi}{\Lambda_{\gamma,n}}\right)^n \frac{F_{\mu\nu}F^{\mu\nu}}{4} - \left(\frac{\phi}{\Lambda_{e,n}}\right)^n m_e \bar{\psi}_e \psi_e - \left(\frac{\phi}{\Lambda_{N,n}}\right)^n m_N \bar{\psi}_N \psi_N
 \, , 
\end{align}
where $F$ is the electromagnetic field tensor, $\psi_e$ denotes the electron field, and $\psi_N$ denote the nucleon fields. 
The parameters $\LL{i}{n}$ denote the effective new-physics energy scales of the underlying model, with $n$ being a positive integer; higher energy scales correspond to feebler interactions between the scalar field and SM fields.
In this section, we employ the natural units $\hbar = c = 1$.

The most commonly considered values of $n$ in Eq.~(\ref{scalar-field_Lagrangian}) are $n=1$ (\textit{linear portal}) and $n=2$ (\textit{quadratic portal}). 
The linear portal is generally devoid of screening effects and linear interactions are absent altogether, e.g., in models with an underlying $Z_2$ symmetry (which precludes interactions with an odd power of $\phi$). 
The quadratic portal is of particular interest to experiments employing at least one atomic clock in space, since 
the scalar field can be strongly screened near the surface of and inside Earth as well as other large dense bodies \cite{Hees:2018screen,Stadnik:2020screen,Stadnik:2021screen,Damour:1993screen,Pospelov:2008screen,Khoury:2010screen}. 
When the scalar field is strongly screened inside Earth, the scalar-field amplitude can be suppressed by the factor of $\sim h/R$ near the surface of a spherical dense body, where $h$ is the height above the surface of the body and $R$ is the radius of the body; for a typical height of a ground-based atomic clock of $h \sim 1~\textrm{m}$, we have $h/R_\oplus \sim 10^{-7}$, which means that the utilisation of space-based clocks, such as in OACESS, can provide an enormous advantage. 

The Lagrangian (\ref{scalar-field_Lagrangian}) induces the following changes in the apparent values of the electromagnetic fine-structure constant $\alpha$ and the fermion masses \cite{Stadnik:2015DM1,Stadnik:2015DM2}: 
\begin{align}
\label{scalar-field_VFCs}
 &\frac{\delta \alpha}{\alpha} \approx \left(\frac{\phi}{\LL{\gamma}{n}}\right)^n,\quad
\frac{\delta m_e}{m_e} = \left(\frac{\phi}{\LL{e}{n}}\right)^n,\quad
\frac{\delta m_N}{m_N} = \left(\frac{\phi}{\LL{N}{n}}\right)^n  \, . 
\end{align}

The response of a clock transition frequency $\nu$ to apparent variations of one or more of the fundamental constants $X = \alpha, m_e, m_N$ can be parameterised in terms of the \textit{relative sensitivity coefficients} $K_X$, according to: 
\begin{equation}
\label{sensitivity_coefficients_definition}
\frac{\delta \nu}{\nu} = \sum_X K_X  \frac{\delta X}{X}  \, . 
\end{equation}

In the non-relativistic limit, an archetypal optical atomic transition frequency scales proportionally to the Rydberg constant, $\nu \propto m_e \alpha^2$, and so $K_\alpha = +2$, $K_{m_e} = +1$ and $K_{m_N} = 0$. 
In frequency comparisons of two \textit{co-located} optical atomic transitions, the Rydberg constant cancels in the frequency ratio and sensitivity to variations of $\alpha$ arises from differences in the relativistic correction factors associated with the two transitions \cite{Flambaum:1999A,Flambaum:1999B}, which have been calculated extensively \cite{Flambaum:1999A,Flambaum:1999B,Flambaum:2003,Flambaum:2004,Flambaum:2006,Flambaum:2008}; e.g., the relativistic factor (in $K_\alpha$ equivalent) is $+0.06$ in the case of the Sr $^1\textrm{S}_0$ $-$ $^3\textrm{P}_0$ optical clock transition, but can have a significantly larger magnitude for transitions in heavier atomic species (e.g., Yb, Hg). 
On the other hand, in ground-to-space comparisons of a \textit{single} optical clock transition, which involve an independent determination of the clock height difference e.g.~via a combination of laser ranging and either orbital position or gravimeter data, the effective sensitivity coefficients are $K_\alpha^\textrm{eff} \approx +2$, $K_{m_e}^\textrm{eff} \approx +1$, $K_{m_N}^\textrm{eff} \approx -1$ in the non-relativistic limit \cite{Stadnik:2020screen}; 
here the sensitivity to $\delta \alpha$ and $\delta m_e$ mainly comes from the Rydberg constant associated with the optical transition, while the sensitivity to $\delta m_N$ mainly comes from the dominant nucleon content of the test masses or bodies. 
Compared to optical atomic transitions, in the non-relativistic limit, an archetypal hyperfine atomic transition has larger relative sensitivity coefficients ($K_\alpha = +4$, $K_{m_e} = +2$ and $K_{m_N} = -1$), due to the short-range nature of the magnetic hyperfine interaction and the participation of the nuclear magnetic moment.

\begin{figure}[ht!]
\includegraphics[width=0.464\linewidth]{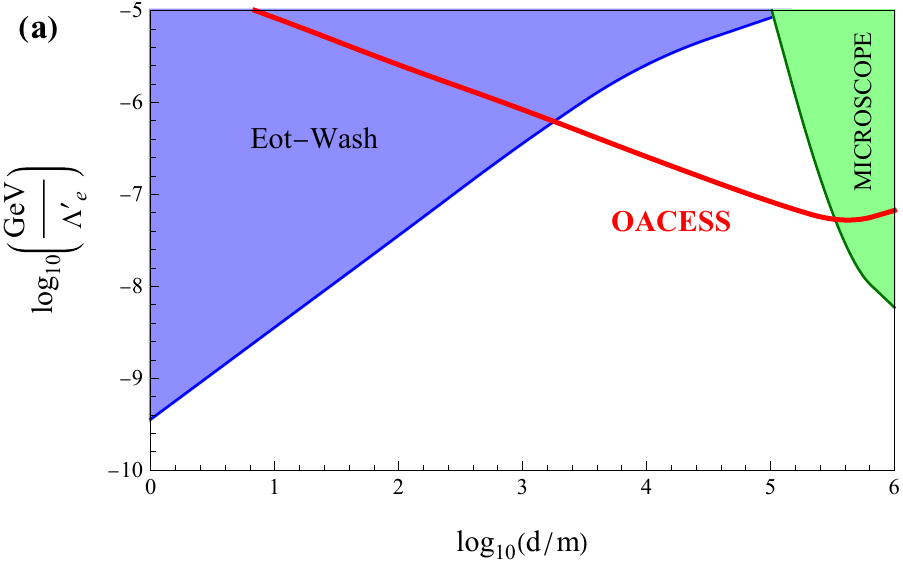}
\hspace{10mm}
\includegraphics[width=0.464\linewidth]{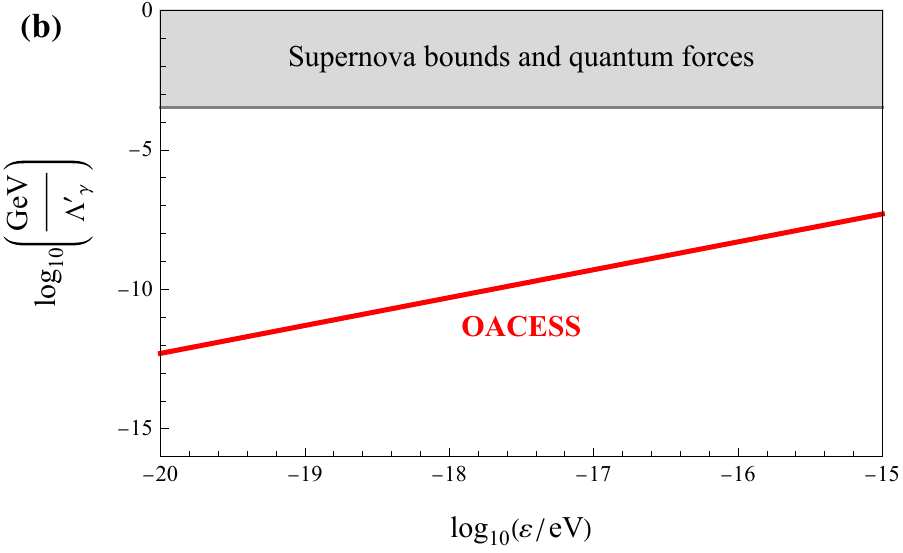}
\caption{
Projected sensitivity of a Sr OACESS clock with a fractional frequency uncertainty of $10^{-18}$ on board the ISS (red curve) to 
\textbf{(a)} static or quasi-static apparent variations of $m_e$ with changing height above Earth's surface, arising in a scalar-field model that permits the production of cosmological domain walls, 
and \textbf{(b)} transient changes in the apparent value of $\alpha$ due to the passage of a relativistic scalar wave from an intense burst of extraterrestrial origin. 
The figures are for the quadratic ($n=2$) interactions of the scalar field with \textbf{(a)} the electron and \textbf{(b)} the electromagnetic field. 
In subfigure \textbf{(a)}, the blue and green regions correspond to torsion-pendulum-type constraints from the ground-based E\"{o}t-Wash experiment and the space-based MICROSCOPE mission, respectively, obtained by rescaling the limits in Ref.~\cite{Stadnik:2020screen} for the model parameters described in the main text of our present paper. 
In subfigure \textbf{(b)}, the grey region corresponds to existing constraints from supernova energy-loss bounds and short-range tests of gravity searching for quantum forces \cite{Pospelov:2008screen}. 
}
\label{fig:projections1}
\end{figure}

In Figs.~\ref{fig:projections1}(a) and \ref{fig:projections1}(b), we present the projected sensitivity of a clock operating in low Earth orbit (e.g., ISS), assuming a fractional frequency stability   of 2$ \times 10^{-16}/\sqrt{\tau/\textrm{s}}$ and a fractional frequency uncertainty of 10$^{-18}$, in the context of two different scalar-field models: 

\textbullet\
In Fig.~\ref{fig:projections1}(a), static or quasi-static apparent variations of $m_e$ with changing height above Earth's surface, arising in a scalar-field model that permits the production of cosmological domain walls \cite{Stadnik:2020screen}. 
Here we have chosen model parameters such that a single cosmological domain wall would contribute to the present-day mass-energy fraction of the Universe at the level of one part in $10^{20}$, with the presence of a single domain wall motivated by results of numerical simulations \cite{Hindmarsh:2003DW,Avelino:2005DW,Martins:2005DW} and the choice of mass-energy fraction consistent with cosmological observations \cite{Stadnik:2020screen,Spergel:1989DW}; however, our sensitivity estimate also equally applies if there are no domain walls of cosmological origin. 
The apparent value of $m_e$ changes over a characteristic height of $\sim \textrm{min} (d,R_\oplus)$ away from Earth's surface, where $d$ is the thickness of a (possible) domain wall of cosmological origin, and so the comparison of the space-based OACESS clock with another Sr clock on the ground can offer an enormous advantage over purely ground-based clock comparisons such as in Ref.~\cite{Takamoto:2020Skytree} for $d \gtrsim 1~\textrm{km}$. 

\textbullet\
In Fig.~\ref{fig:projections1}(b), transient changes in the apparent value of $\alpha$ due to the passage of a relativistic scalar wave from an intense burst of extraterrestrial origin \cite{Stadnik:2021screen,Dailey:2021multi}. 
Here we have chosen burst parameters that yield a coherent burst over the entire relevant range of scalar particle energies $\varepsilon$ and have assumed that the same type of scalar particle makes no contribution towards the matter-energy content of the Universe (beyond the negligible fraction due to the bursts themselves). 
The use of a space-based clock network with multiple nodes would allow for the determination of the direction to the burst source. 

In both cases, analogous ground-based clock experiments have insufficient sensitivity to probe the relevant parameter space in these models due to strong screening of the scalar field by Earth's atmosphere (see Refs.~\cite{Stadnik:2020screen} and \cite{Stadnik:2021screen} for details). 
Experiments utilising space-based optical clocks therefore open up possibilities that are currently inaccessible to purely ground-based experiments.

\section{Space Station Enabled Optical Clock Hardware}

\begin{figure}[tb!]
\centering
\includegraphics[width=16.25cm]{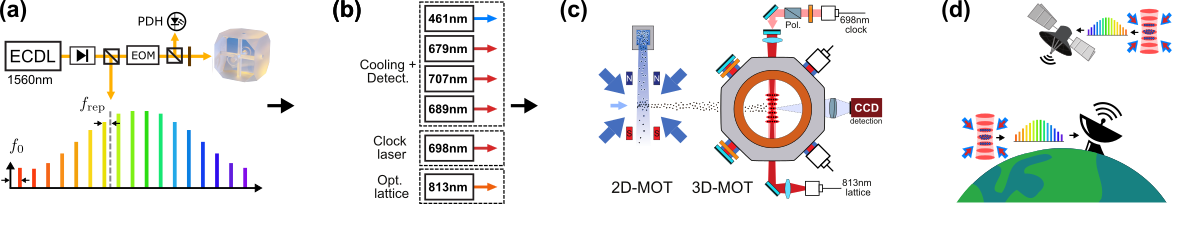}
\caption{Schematic representation of the payload. 
\textbf{(a)} A compact and high stability clock laser will be stabilized via a high-finesse optical cavity for short-term stability and to the atoms for long-term stability.  A frequency comb locked to the clock laser transfers the stability between the IR region and the clock wavelength at 698\,nm. 
\textbf{(b)} The diode laser system in OACESS will be based on technologies operated in sounding rocket missions which are currently being developed further for space applications. 
\textbf{(c)} The science module is matured from CAL and JPL heritage to provide high atom flux of ultra cold strontium gases with low SWaP. 
\textbf{(d)} O-TWTFT will allow precise space-to-ground comparisons of high-precision optical clocks worldwide.}
\label{fig5}
\end{figure}

Key components for a space-based optical atomic clock payload include the science module (in which the atoms are produced, cooled, and interrogated), the laser and optical system (including the ultra-high-stability clock laser), and the clock-signal distribution system (including the optical frequency comb (OFC) for down-converting high-precision clock signals from the optical domain and space-to-ground comparison via an optical link). The payload is illustrated in Fig.~\ref{fig5}.  In the following, we briefly discuss the primary components for OACESS and the ongoing efforts from various institutions worldwide towards supporting a near-term Earth-orbiting optical clock mission. 

\subsection*{Science Module}

OACESS will require a rugged science module that incorporates the physics package (vacuum chamber), magnetic coils, optomechanics, magnetic shields, and thermal control systems in a low-SWaP design. The physics package will consist of a low-power, high-flux atom source and an ultra-high-vacuum (UHV) (better than $10^{-10}$ torr) ``science cell'' with exceptional optical access to accommodate three retro-reflected pairs of laser beams for a two-stage Magneto-Optical Trap (MOT), an 813\,nm lattice laser, a clock laser, optical pumping beams, and an independent imaging path. 
Mature solutions currently exist for the miniaturized Sr source \cite{AOSENSE,Schioppo:2012}. 

We envision that the science cell, vacuum pumps, magnetic coils, opto-mechanics, magnetic shields, cooling, electronics, and control/imaging systems can all strongly leverage the heritage established from NASA's Cold Atom Lab (CAL) \cite{Aveline2020} as well as ongoing DLR activities (SOLIS-1G). The latter aims to develop a compact, low-SWaP 2D-MOT atomic source and a 3D-MOT based on an aluminum alloy with maximized optical access. The non-magnetic alloy has a low level of helium permeability, allowing to reach low UHV levels without high temperature baking and high pumping speed pumps. Adaptations that might be beneficial for OACESS, like a pyramidal single beam 3D-MOT assembly or an integrated enhancement cavity for the optical lattice will be investigated.

However, because some types of OACESS dark-sector searches rely on time-dependent comparisons of ultra-high-precision clocks, it will require unprecedented environmental control with respect to space-based quantum sensors. Influences and mitigation plans for the leading clock systematic errors will therefore be an integral part of this study. The development and use of ground testbeds for analyzing/testing OACESS designs for functionality and unanticipated systematic shifts will be critical.

\subsection*{Laser and Optical System}
The operation of OACESS requires a laser system capable of two-stage laser cooling, state preparation, coherent transfer of atoms and trapping in an optical lattice at the magic wavelength, and ultra-stable interrogation of the clock transition \cite {RevModPhys.87.637}. The required wavelengths are shown in Figure \ref{fig5}. Here, the ultra-stable clock laser system will combine mature cavity-stabilized laser designs from terrestrial labs  \cite {RevModPhys.87.637,Campbell2017,McGrew2018,Schioppo2017,Oelker2019,Hinkley2013,Marti2018,Bothwell2019} with flight-qualified laser system designs (e.g. for the GRACE-FO mission \cite{Thompson:2011}).

Strontium optical clock technologies have matured to the point where all necessary wavelengths can be provided by diode lasers. Diode-laser-based systems already operate in experiments at the Bremen drop tower, on sounding rockets, and in CAL to study ultra-cold rubidium and potassium atoms \cite{Schiemangk:15,Pahl:19, Becker:2018, Aveline2020}. Laser payloads and optical frequency references have been developed for operation on sounding rockets and reached TRL 9 on such sub-orbital vehicles through flight mission operation \cite{Kohfeldt:2016,Luvsandamdin:2014,Lezius2016,Dinkelaker:2017}.
These payloads feature diode laser systems which are based on a micro-integrated laser technology platform, providing compact, robust and energy-efficient semiconductor laser modules. Moreover, a system for laser cooling and atom interferometry with $^{87}$Rb atoms was launched in 2017 (MAIUS) \cite{Schkolnik:2016} and a high-performance optical frequency reference based on molecular iodine (JOKARUS mission) has also been launched \cite{Doeringshoff:19, Schkolnik:2017}. Currently, laser systems with even lower SWaP are under development for future operation on CubeSats \cite{Strangfeld:2020}, integrated into standalone piggyback devices \cite{Strangfeld:2022} where flexible field-programmable-gate-array-based tools are used for automatic lock point selection and autonomously optimising spectroscopy parameters \cite{Wiegand:2022}. 

As part of the ongoing DLR SOLIS-1G activities, DBR lasers optimized for  strontium lattice clocks, at 689.449\,nm were fabricated. The devices achieve more than 80\,mW as well as a spectral linewidth of 0.4\,MHz \cite{Pyrlik:2022}. Extending the wavelength range from 679\,nm to 707\,nm would allow for a highly integrated and intrinsically stable laser system. A space qualified frequency comb (Menlo Systems) as well as SMD-based light modulators (QUBIG) are being developed or first prototypes exist.

Development of a robust, low SWaP clock laser system will benefit from early development of ground test-beds to validate the design and performance of the optical clock.

\subsection*{Frequency comparison}

We consider frequency-comb-based optical two-way time-frequency transfer (O-TWTFT) \cite{BACON2021,Bodine2020} as one highly promising method for achieving space-to-ground clock comparisons below the $10^{-18}$ level. Here, laser light stabilized by the optical clocks would be transmitted to optical frequency combs. OFC pulses from Earth-based clocks and from the space-based clock would then be simultaneously exchanged via a free-space optical link.
This approach mirrors conventional RF two-way satellite transfer but achieves much higher precision by use of short optical pulses. In this two-way approach, the time-of-flight fluctuations are canceled to yield just the relative clock offsets. Recent demonstrations have compared state-of-the-art optical clocks at JILA and NIST, where the residual instability contribution from the O-TWTFT comparison was well below $10^{-18}$ \cite{BACON2021,Bodine2020}. O-TWTFT has also been demonstrated over a link to a flying quad-copter at closing velocities of up to 20\,m/s with no observable degradation from velocity-dependent systematics \cite{Bergeron:2019}.

Frequency combs have already been operated on suborbital vehicles and optical frequency measurements were successfully demonstrated on two sounding rocket flights \cite{Lezius2016} and are currently being developed for operation on the Bartolomeo platform attached to  the ISS (COMPASSO). Using fiber amplifiers and second-harmonic generation, the spectrum from 400\,nm - 2000\,nm can be covered. 
In this way, all lasers required for clock operation could be referenced directly to the comb, allowing for additional stabilized laser sources for high-performance space-based optical-to-ground links.  For this mission, mature designs for a robust, low-SWaP frequency comb will need to be space qualified. In addition, the promising O-TWTFT transfer scheme needs to be validated for orders of magnitude longer length scales and Doppler shifts for the ground-to-space OAC comparisons.

A modified version of the O-TWTFT technique shows promise for operation at significantly lower received powers \cite{Caldwell:2022}, which should enable operation at low power-aperture products over the long path lengths between a clock in low-earth orbit and on the ground.  However, this new approach will require further development to operate at the high closing velocities associated with low-earth-orbit satellites.

\section{Summary}
 OACESS aims to serve as a pathfinder science mission to search for space-time-dependent signatures of dark scalar fields while establishing the technologies for high-precision optical atomic clocks in space. This mission will directly leverage the developments and lessons learned from the recent and ongoing campaigns of NASA's CAL, several DLR sounding rocket missions that demonstrated laser systems, BEC-based interferometery, optical atomic and molecular frequency references, as well as future missions under development like the NASA/DLR BECCAL. As summarized in the BPS Topical White paper ``Space-Time Referencing:~atomic clocks, laser links and applications'' \cite{Hollberg2022}, space-based clocks and Earth-space timing links,  building off of the technologies matured by OACESS, can provide enabling capabilities for geodesy, Earth sciences, navigation, world-wide time transfer, and fundamental physics including testing the foundations of General Relativity and searching for physics beyond the Standard Model.

\subsection*{Acknowledgments}

\textbf{V.S.}, \textbf{O.F.}, \textbf{M.K.}, \textbf{C.P.} and \textbf{I.T.}. acknowledge support by the German Space Agency (DLR) with funds provided by the Federal Ministry of Economic Affairs and Climate Action (BMWK) under grant numbers DLR 50WM 2151, 2052 and by the German Federal Ministry of Education and Research within the program quantum technologies - from basic research to market under grant number 13N15725.

The work of \textbf{D.B.} was supported in part by the Deutsche Forschungsgemeinschaft (DFG, German Research Foundation) Project ID 390831469:  EXC 2118 (PRISMA+ Cluster of Excellence).

\textbf{V.F.}'s work was supported by Australian Research Council Grants No. DP190100974 and DP200100150.

\textbf{L.H.} acknowledges partial support from: the NASA BPS Fundamental Physics Program and the Office of Naval Research Grant No. N000141712255.

\textbf{S.K.}’s work was supported by the NIST Precision Measurement Grants program, the Northwestern University Center for Fundamental Physics and the John Templeton Foundation through a Fundamental Physics grant, a Packard Fellowship for Science and Engineering, the Army Research Office through agreement number W911NF-21-1-0012, and the National Science Foundation under Grant No. 2143870.

\textbf{A.L.} acknowledge support from NIST, ONR, and NSF QLCI.

\textbf{N.R.N.} and \textbf{L.S.} acknowledge support from the NASA BPS Fundamental Physics Program and NIST.

The work of \textbf{Y.V.S.} was supported by the World Premier International Research Center Initiative (WPI), MEXT, Japan and by the JSPS KAKENHI Grant Number JP20K14460, as well as by the Australian Research Council under the Discovery Early Career Researcher Award DE210101593.

\textbf{J.Y.} acknowledge support from the National Science Foundation QLCI OMA-2016244, the DOE Quantum System Accelerator, and NIST. 

\textbf{J.R.W.} is supported by the National Aeronautics and Space Administration through a contract with the Jet Propulsion Laboratory, California Institute of Technology.



\begin{thebibliography}{99}




\bibitem{Safronova2018}
M. Safronova \textit{et al}., 
Search for new physics with atoms and molecules,
\href{https://journals.aps.org/rmp/abstract/10.1103/RevModPhys.90.025008}{Rev. Mod. Phys. \textbf{90}, 025008 (2018)}. 

\bibitem{antypas2022new}
D. Antypas \textit{et al}., 
New Horizons: Scalar and Vector Ultralight Dark Matter,
\href{https://arxiv.org/pdf/2203.14915.pdf}{arXiv preprint arXiv:2203.14915 (2022)}.

\bibitem{Bothwell2022}
T. Bothwell \textit{et al}., 
Resolving the gravitational redshift across a millimetre-scale atomic sample,
\href{https://www.nature.com/articles/s41586-021-04349-7}{Nature ~\textbf{602}, 420 (2022)}.

\bibitem{Zheng2022}
X. Zheng \textit{et al}., 
Differential clock comparisons with a multiplexed optical lattice clock,
\href{https://www.nature.com/articles/s41586-021-04344-y#article-info}{Nature ~\textbf{602}, 425 (2022)}.

\bibitem{Matei2017}
D. Matei \textit{et al}., 
1.5 $\mu$m lasers with Sub-10 mHz Linewidth,
\href{https://journals.aps.org/prl/abstract/10.1103/PhysRevLett.118.263202}{Phys. Rev. Lett ~\textbf{118}, 263202 (2017)}.

\bibitem{Zhang2017}
W. Zhang \textit{et al}., 
Ultrastable Silicon Cavity in a Continuously Operating Closed-Cycle Cryostat at 4 K,
\href{https://journals.aps.org/prl/abstract/10.1103/PhysRevLett.119.243601}{Phys. Rev. Lett ~\textbf{119}, 243601 (2017)}.

\bibitem{RevModPhys.87.637}
A. Ludlow \textit{et al}., 
Optical atomic clocks,
\href{https://journals.aps.org/rmp/abstract/10.1103/RevModPhys.87.637}{Rev. Mod. Phys. ~\textbf{87}, 637 (2015)}. 

\bibitem{Campbell2017}
S. L. Campbell \textit{et al}., 
A Fermi-degenerate three-dimensional optical lattice clock,
\href{https://www.science.org/doi/10.1126/science.aam5538}{Science ~\textbf{358}, 6359 (2017)}. 

\bibitem{McGrew2018}
 W. F. McGrew \textit{et al}., 
Atomic clock performance enabling geodesy below the centimetre level,
\href{https://www.nature.com/articles/s41586-018-0738-2}{Nature ~\textbf{564}, 87-90 (2018)}. 

\bibitem{Schioppo2017}
M. Schioppo \textit{et al}., 
Ultrastable optical clock with two cold-atom ensembles,
\href{https://www.nature.com/articles/nphoton.2016.231}{Nat. Photon. ~\textbf{11}, 48-52 (2017)}. 

\bibitem{Oelker2019}
E. Oelker \textit{et al}., 
Demonstration of $4.8 \times 10^{-17}$ stability at 1 s for two independent optical clocks,
\href{https://www.nature.com/articles/s41566-019-0493-4}{Nat. Photon. ~\textbf{13}, 714-719 (2019)}. 
 
\bibitem{Hinkley2013}
N. Hinkley \textit{et al}., 
An atomic clock with $10^{-18}$ instability,
\href{https://www.science.org/doi/10.1126/science.1240420}{Science ~\textbf{341}, 1215-1218 (2013)}. 
 
 \bibitem{Marti2018}
G. E. Marti \textit{et al}., 
Imaging optical frequencies with 100 µHz precision and 1.1 µm resolution,
\href{https://journals.aps.org/prl/abstract/10.1103/PhysRevLett.120.103201}{Phys. Rev. Lett. ~\textbf{120}, 103201 (2018)}. 
 
 \bibitem{Bothwell2019}
T. Bothwell \textit{et al}., 
JILA SrI optical lattice clock with uncertainty of $2.0 \times 10^{-18}$,
\href{https://iopscience.iop.org/article/10.1088/1681-7575/ab4089}{Metrologia ~\textbf{56}, 065004 (2019)}. 

 

\bibitem{BACON2021}
Boulder Atomic Clock Optical Network (BACON) Collaboration,
Frequency ratio measurements at 18-digit accuracy using an optical clock network,
\href{https://doi.org/10.1038/s41586-021-03253-4}{Nature ~\textbf{591}, 564 (2021)}.

\bibitem{Komar2104}
P. K\'{o}m\'{a}r \textit{et al}., 
A quantum network of clocks,
\href{https://www.nature.com/articles/nphys3000}{Nature Phys ~\textbf{10}, 582 (2014)}.

\bibitem{Schuldt:2021}
T.~Schuldt \textit{et al}., 
Optical clock technologies for global navigation satellite systems,
\href{https://doi.org/10.1007/s10291-021-01113-2}{GPS Solut. \textbf{25},  83  (2021)}.  
  
 

\bibitem{Hees:2018screen}
A.~Hees, O.~Minazzoli, E.~Savalle, Y.~V.~Stadnik, and P.~Wolf, 
Violation of the equivalence principle from light scalar dark matter, 
\href{https://journals.aps.org/prd/abstract/10.1103/PhysRevD.98.064051}{Phys.~Rev.~D \textbf{98}, 064051 (2018)}. 

\bibitem{Stadnik:2020screen}
Y.~V.~Stadnik, 
New bounds on macroscopic scalar-field topological defects from nontransient signatures due to environmental dependence and spatial variations of the fundamental constants, 
\href{https://journals.aps.org/prd/abstract/10.1103/PhysRevD.102.115016}{Phys.~Rev.~D \textbf{102}, 115016 (2020)}. 

\bibitem{Stadnik:2021screen}
Y.~V.~Stadnik, 
Comment on ``Quantum sensor networks as exotic field telescopes for multi-messenger astronomy'', 
\href{https://arxiv.org/abs/2111.14351}{arXiv:2111.14351}. 


\bibitem{Damour:1993screen}
T.~Damour and G.~Esposito-Farese, 
Nonperturbative strong-field effects in tensor-scalar theories of gravitation, 
\href{https://journals.aps.org/prl/abstract/10.1103/PhysRevLett.70.2220}{Phys.~Rev.~Lett.~\textbf{70}, 2220 (1993)}. 

\bibitem{Pospelov:2008screen}
K.~A.~Olive and M.~Pospelov, 
Environmental dependence of masses and coupling constants, 
\href{https://journals.aps.org/prd/abstract/10.1103/PhysRevD.77.043524}{Phys.~Rev.~D \textbf{77}, 043524 (2008)}. 

\bibitem{Khoury:2010screen}
K.~Hinterbichler and J.~Khoury, 
Screening Long-Range Forces through Local Symmetry Restoration, 
\href{https://journals.aps.org/prl/abstract/10.1103/PhysRevLett.104.231301}{Phys.~Rev.~Lett.~\textbf{104}, 231301 (2010)}. 

 \bibitem{Decadal2021}
 R.~Feryl and K.~Van Vliet, 
\href{file:///Users/jrwillia/Downloads/Call%20Letter%20with%20Specifications%20(4).pdf}{Call to the Biological and Physical Sciences in Space Community for White Papers (2021)}.

 \bibitem{Bodine2020}
M.~I.~Bodine \textit{et al}., 
Optical atomic clock comparison through turbulent air,
\href{https://jila.colorado.edu/sites/default/files/2020-10/PhysRevResearch.2.033395.pdf}{Phys.~Rev.~Res.~\textbf{2}, 033395 (2020)}. 

 \bibitem{Hollberg2022}
L.~Hollberg \textit{et al}., Space-Time Referencing:~atomic clocks, laser links and applications,
\href{https://www.nationalacademies.org/our-work/decadal-survey-on-life-and-physical-sciences-research-in-space-2023-2032}{A Topic White Paper submitted to the NASA Decadal Survey on Biological and Physical Sciences Research in Space 2023-2032, Dec. (2021)}. 

 \bibitem{Berceau2016}
P.~Berceau \textit{et al}., 
Space-time reference with an optical link.,
\href{https://ee.stanford.edu/~jmk/pubs/STR.CQG.2016.pdf}{Classical and Quantum Gravity 33.13 (2016)}. 

 \bibitem{Robert2016}
C.~Robert, J.-M.~Conan, and P.~Wolf, Impact of turbulence on high-precision ground-satellite frequency transfer with two-way coherent optical links, \href{https://journals.aps.org/pra/abstract/10.1103/PhysRevA.93.033860}{Phys. Rev. A, 93, 033860 (2016)}.

 \bibitem{Taylor2017}
M.~T.~Taylor, L.~Hollberg, J.~M.~Kahn, Effect of Atmospheric Anisoplanatism on Earth-to-Satellite Time Transfer over Laser Communication Links, \href{https://doi.org/10.1364/OE.25.015676}{Opt. Express, 25, 14, 15676–15686 (2017)}. 

 \bibitem{Taylor2020}
M.~T.~Taylor, \textit{et al}., Effect of Atmospheric Turbulence on Timing Instability for Partially Reciprocal Two-Way Optical Time Transfer Links, \href{https://doi.org/10.1103/PhysRevA.101.033843}{Phys. Rev. A, 101, 3, 033843 (2020)}.

\bibitem{FOCOS2021}
A.~Derevianko \textit{et al}., 
Fundamental Physics with a State-of-the-Art Optical Clock in Space,
\href{https://doi.org/10.1088/2058-9565/ac7df9}{Quantum Science and Technology,  accepted, (2022)}. 

\bibitem{Acuna2002}
M.~H.~Acu\~{n}a, Space-based magnetometers, \href{https://corpora.tika.apache.org/base/docs/govdocs1/514/514160.pdf}{Rev. Sci. Inst. 73, 11, 3717, (2002)}.

\bibitem{Korth2016} 
H.~Korth, \textit{et al}., Miniature atomic scalar magnetometer for space based on the rubidium isotope 87Rb, \href{https://agupubs.onlinelibrary.wiley.com/doi/full/10.1002/2016JA022389}{J. Geophys. Res. Space Physics, 121, 7870–7880 (2016)}. 



\bibitem{Arvanitaki:2015DM}
A.~Arvanitaki, J.~Huang and K.~Van Tilburg, Searching for dilaton dark matter with atomic clocks, 
\href{https://journals.aps.org/prd/abstract/10.1103/PhysRevD.91.015015}{Phys.~Rev.~D \textbf{91}, 015015 (2015)}. 

\bibitem{Stadnik:2015DM1}
Y.~V.~Stadnik and V.~V.~Flambaum, 
Searching for Dark Matter and Variation of Fundamental Constants with Laser and Maser Interferometry, 
\href{https://journals.aps.org/prl/abstract/10.1103/PhysRevLett.114.161301}{Phys.~Rev.~Lett.~\textbf{114}, 161301 (2015)}. 

\bibitem{Stadnik:2015DM2}
Y.~V.~Stadnik and V.~V.~Flambaum,
Can dark matter induce cosmological evolution of the fundamental constants of Nature?, 
\href{https://journals.aps.org/prl/abstract/10.1103/PhysRevLett.115.201301}{Phys.~Rev.~Lett.~\textbf{115}, 201301 (2015)}. 


\bibitem{Derevianko:2014solitons}
A.~Derevianko and M.~Pospelov, 
Hunting for topological dark matter with atomic clocks, 
\href{https://www.nature.com/articles/nphys3137}{Nat.~Phys.~\textbf{10}, 933 (2014)}. 


\bibitem{Dailey:2021multi}
C.~Dailey \textit{et al}., 
Quantum sensor networks as exotic field telescopes for multi-messenger astronomy, 
\href{https://www.nature.com/articles/s41550-020-01242-7}{Nat.~Astron.~\textbf{5}, 150 (2021)}. 




  

\bibitem{Flambaum:1999A}
V.~A.~Dzuba, V.~V.~Flambaum, and J.~K.~Webb, 
Space-Time Variation of Physical Constants and Relativistic Corrections in Atoms, 
\href{https://journals.aps.org/prl/abstract/10.1103/PhysRevLett.82.888}{Phys.~Rev.~Lett.~\textbf{82}, 888 (1999)}. 

\bibitem{Flambaum:1999B}
V.~A.~Dzuba, V.~V.~Flambaum, and J.~K.~Webb, 
Calculations of the relativistic effects in many-electron atoms and space-time variation of fundamental constants, 
\href{https://journals.aps.org/pra/abstract/10.1103/PhysRevA.59.230}{Phys.~Rev.~A \textbf{59}, 230 (1999)}. 

\bibitem{Flambaum:2003}
V.~A.~Dzuba, V.~V.~Flambaum, and M.~V.~Marchenko, 
Relativistic effects in Sr, Dy, Yb II, and Yb III and search for variation of the fine-structure constant, 
\href{https://journals.aps.org/pra/abstract/10.1103/PhysRevA.68.022506}{Phys.~Rev.~A \textbf{68}, 022506 (2003)}. 

\bibitem{Flambaum:2004}
E.~J.~Angstmann, V.~A.~Dzuba, and V.~V.~Flambaum, 
Relativistic effects in two valence-electron atoms and ions and the search for variation of the fine-structure constant, 
\href{https://journals.aps.org/pra/abstract/10.1103/PhysRevA.70.014102}{Phys.~Rev.~A \textbf{70}, 014102 (2004)}. 

\bibitem{Flambaum:2006}
V.~V.~Flambaum and A.~F.~Tedesco, 
Dependence of nuclear magnetic moments on quark masses and limits on temporal variation of fundamental constants from atomic clock experiments, 
\href{https://journals.aps.org/prc/abstract/10.1103/PhysRevC.73.055501}{Phys.~Rev.~C \textbf{73}, 055501 (2006)}. 

\bibitem{Flambaum:2008}
V.~A.~Dzuba and V.~V.~Flambaum, 
Relativistic corrections to transition frequencies of Ag I, Dy I, Ho I, Yb II, Yb III, Au I, and Hg II and search for variation of the fine-structure constant, 
\href{https://journals.aps.org/pra/abstract/10.1103/PhysRevA.77.012515}{Phys.~Rev.~A \textbf{77}, 012515 (2008)}. 

\bibitem{Hindmarsh:2003DW} T.~Garagounis and M.~Hindmarsh, Scaling in numerical simulations of domain walls, \href{https://journals.aps.org/prd/abstract/10.1103/PhysRevD.68.103506}{Phys.~Rev.~D \textbf{68}, 103506 (2003)}. 

\bibitem{Avelino:2005DW} J.~C.~R.~E.~Oliveira, C.~J.~A.~P.~Martins, and P.~P.~Avelino, The cosmological evolution of domain wall networks, \href{https://journals.aps.org/prd/abstract/10.1103/PhysRevD.71.083509}{Phys.~Rev.~D \textbf{71}, 083509 (2005)}. 

\bibitem{Martins:2005DW} P.~P.~Avelino, J.~C.~R.~E.~Oliveira, and C.~J.~A.~P.~Martins, Understanding domain wall network evolution, \href{https://www.sciencedirect.com/science/article/pii/S0370269305001814?via\%3Dihub}{Phys.~Lett.~B \textbf{610}, 1 (2005)}. 

\bibitem{Spergel:1989DW} W.~H.~Press, B.~S.~Ryden, and D.~N.~Spergel, Dynamical evolution of domain walls in an expanding Universe, \href{https://ui.adsabs.harvard.edu/abs/1989ApJ...347..590P/abstract}{Astrophys.~J.~\textbf{347}, 590 (1989)}. 

\bibitem{Takamoto:2020Skytree}
M.~Takamoto \textit{et al}., 
Test of general relativity by a pair of transportable optical lattice clocks, 
\href{https://www.nature.com/articles/s41566-020-0619-8}{Nat.~Photonics \textbf{14}, 411 (2020)}. 
  
\bibitem{AOSENSE}
\href{https://aosense.com/product/cold-atomic-beam-system-reload/}{https://aosense.com/product/cold-atomic-beam-system-reload/}  

\bibitem{Schioppo:2012}
M.~Schioppo \textit{et al}., 
A compact and efficient strontium oven for laser-cooling experiments, 
\href{https://aip.scitation.org/doi/10.1063/1.4756936}{Rev. Sci. Inst \textbf{83}, 103101 (2012)}.   

\bibitem{Aveline2020}
D.~Aveline \textit{et al}., 
Observation of Bose–Einstein condensates in an Earth-orbiting research lab, 
\href{https://www.nature.com/articles/s41586-020-2346-1}{Nature \textbf{582}, 193-197 (2020)}. 

\bibitem{Thompson:2011}
R.~Thompson \textit{et al}., 
A flight-like optical reference cavity for GRACE follow-on laser frequency stabilization, 
\href{https://ieeexplore.ieee.org/document/5977873}{Joint Conference of the IEEE International Frequency Control and the European Frequency and Time Forum (FCS) Proceedings, 1-2 (2011)}.  


\bibitem{Schiemangk:15}
M.~Schiemangk \textit{et al}., 
High-power, micro-integrated diode laser modules at 767 and 780 nm for portable quantum gas experiments, 
\href{https://www.osapublishing.org/ao/abstract.cfm?uri=ao-54-17-5332}{Appl. Opt. \textbf{54}, 5332-5338 (2015)}. 
  
\bibitem{Pahl:19}
J.~Pahl \textit{et al}., 
Compact and robust diode laser system technology for dual-species ultracold atom experiments with rubidium and potassium in microgravity, 
\href{https://doi.org/10.1364/AO.58.005456}{Applied Optics Vol. 58, Issue 20, pp. 5456-5464 (2019)}.

\bibitem{Becker:2018}
D.~Becker \textit{et al}., 
Space-borne Bose–Einstein condensation for precision interferometry,
\href{https://doi.org/10.1038/s41586-018-0605-1}{Nature 562, 391–395 (2018)}.  
    

 \bibitem{Kohfeldt:2016}
A.~Kohfeldt \textit{et al}., 
Compact narrow linewidth diode laser modules for precision quantum optics experiments on board of sounding rockets, 
\href{https://doi.org/10.1117/12.2231537}{SPIE Phot. Europe 2016  pp. 203-212 (2016)}.
  
  
  
\bibitem{Luvsandamdin:2014}
E.~Luvsandamdin \textit{et al}., 
Micro-integrated extended cavity diode lasers for precision potassium spectroscopy in space, 
\href{https://www.osapublishing.org/oe/fulltext.cfm?uri=oe-22-7-7790&id=282284}{Opt. Express  \textbf{22}, pp.  7790-7798 (2014)}.
  
\bibitem{Lezius2016}
M.~Lezius \textit{et al}., 
Space-borne frequency comb metrology, 
\href{https://www.osapublishing.org/optica/fulltext.cfm?uri=optica-3-12-1381&id=354726}{Optica \textbf{12}, pp. 1381-1387 (2016)}.   

\bibitem{Dinkelaker:2017}
A.~Dinkelaker \textit{et al}., 
Autonomous frequency stabilization of two extended-cavity diode lasers at the potassium wavelength on a sounding rocket, 
\href{https://www.osapublishing.org/ao/abstract.cfm?uri=ao-56-5-1388}{Appl. Opt. \textbf{56}, pp. 1388-1396 (2017)}.
    
\bibitem{Schkolnik:2016}
V.~Schkolnik \textit{et al}., 
A compact and robust diode laser system for atom interferometry on a sounding rocket, 
\href{https://link.springer.com/article/10.1007/s00340-016-6490-0}{Appl. Phys. B. \textbf{122}, 217 (2016)}.
  
\bibitem{Doeringshoff:19}
K.~Doeringshoff \textit{et al}., 
Iodine Frequency Reference on a Sounding Rocket, 
\href{https://journals.aps.org/prapplied/abstract/10.1103/PhysRevApplied.11.054068}{ Phys. Rev. Applied 11, 054068 (2019)}.  

\bibitem{Schkolnik:2017}
V.~Schkolnik \textit{et al}., 
 JOKARUS - design of a compact optical iodine frequency reference for a sounding rocket mission, 
\href{https://epjquantumtechnology.springeropen.com/articles/10.1140/epjqt/s40507-017-0063-y}{EPJ Quant. Tech. \textbf{4}, 9 (2017)}.
 
 \bibitem{Strangfeld:2020}
A.~Strangfeld \textit{et al}., 
Prototype of a compact rubidium-based optical frequency reference for operation on nanosatellites, 
\href{https://doi.org/10.1364/JOSAB.420875}{Journal of the Optical Society of America B Vol. 38, Issue 6, pp. 1885-1891 (2021)}.  

\bibitem{Strangfeld:2022}
A.~Strangfeld, B.~Wiegand, J.~Kluge, M.~Schoch, and M.~Krutzik, Compact plug and play optical frequency reference device based on Doppler-free spectroscopy of rubidium vapor, \href{https://opg.optica.org/oe/fulltext.cfm?uri=oe-30-7-12039&id=470786}{Opt.~Express \textbf{30}, 12039 (2022)}. 

\bibitem{Wiegand:2022}
B.~Wiegand, B.~Leykauf, R.~Jördens, and M.~Krutzik, Linien:~A versatile, user-friendly, open-source FPGA-based tool for frequency stabilization and spectroscopy parameter optimization, \href{https://aip.scitation.org/doi/full/10.1063/5.0090384}{Rev.~Sci.~Instrum.~\textbf{93}, 063001 (2022)}. 
   
\bibitem{Pyrlik:2022}
C.~Pyrlik , N.~Goossen-Schmidt , M.~T.~Hassan , A.~Bawamia, J.~Fricke ,
A.~Knigge , A.~Maaßdorf, M.~Schiemangk, H.~Wenzel , and A.~Wicht, High Power Distributed Bragg Reflector Lasers at 689.45\, nm for Quantum Technology Applications, \href{https://doi.org/10.1109/LPT.2021.3139433}{IEEE Phot.~ Tech.~ Letters ~\textbf{34}, 13, pp. 679-682 (2022)}. 

\bibitem{Bergeron:2019}
H.~Bergeron \textit{et al}., 
Femtosecond time synchronization of optical clocks off of a flying quadcopter, 
\href{https://www.nature.com/articles/s41467-019-09768-9}{Nat. Comm. \textbf{10}, 1819  (2019)}.  
  
\bibitem{Schuldt:2015}
T.~Schuldt \textit{et al}., 
Design of a dual species atom interferometer for space, 
\href{https://doi.org/10.1007/s10686-014-9433-y}{Exp. Astron. \textbf{39},  pp. 167–206  (2015)}.  
 
 

 
\bibitem{Oi:2017}
D.~Oi \textit{et al}., 
Nanosatellites for quantum science and technology, 
\href{https://www.tandfonline.com/doi/full/10.1080/00107514.2016.1235150}{Cont. Physs. \textbf{58}, pp. 25-52  (2017)}.  
 
    

\bibitem{Caldwell:2022}
E.~Caldwell, L.~Sinclair, W.~Swann, N.~Newbury, B.~Stuhl, J.~ Deschenes,
Photon Efficient Optical Time Transfer, \href{https://eftf-ifcs2022.sciencesconf.org/}{EFTF-IFCS, Timekeeping, T \& F Transfer, Telecom, GNSS and Applications \textbf{5036},  (2022)}. 

\end{thebibliography}
\end{document}